\documentclass[11pt]{article}

\usepackage[margin=1in]{geometry}
\usepackage{amsmath,amssymb,amsfonts,amsthm,mathtools,bm}
\usepackage{booktabs}
\usepackage{float}
\usepackage{cite}
\usepackage{enumitem}
\usepackage{microtype}
\usepackage{tikz}
\usetikzlibrary{arrows.meta,positioning,fit}
\usepackage[hidelinks]{hyperref}

\newtheorem{theorem}{Theorem}
\newtheorem{proposition}{Proposition}

\DeclareMathOperator{\rank}{rank}
\DeclareMathOperator{\range}{range}

\newcommand{\Herm}[1]{\mathcal H\!\left(#1\right)}
\newcommand{\Id}{\mathbf I}
\newcommand{\Evec}{\mathbf E}
\newcommand{\Vvec}{\mathbf V}
\newcommand{\Ivec}{\mathbf I}
\newcommand{\avec}{\mathbf a}
\newcommand{\bvec}{\mathbf b}
\newcommand{\Smat}{\mathbf S}
\newcommand{\Zload}{\mathbf Z}
\newcommand{\Zs}{\mathbf Z_{\mathrm s}}
\newcommand{\Rs}{\mathbf R_{\mathrm s}}
\newcommand{\Rload}{\mathbf R}
\newcommand{\Rrad}{\mathbf R_{\mathrm{rad}}}
\newcommand{\Rloss}{\mathbf R_{\Omega}}
\newcommand{\TARC}{\operatorname{TARC}}
\newcommand{\Aacc}{\mathbf A_{\mathrm{acc}}}
\newcommand{\Arad}{\mathbf A_{\mathrm{rad}}}
\newcommand{\Aloss}{\mathbf A_{\Omega}}

\title{Maximum-Power-Transfer Power Coordinates\\
for Fully Coupled Multiport Th\'evenin Sources}

\author{Majid Manteghi\\
\texttt{manteghi@vt.edu}}

\date{}

\hypersetup{
  pdftitle={Maximum-Power-Transfer Power Coordinates for Fully Coupled Multiport Thevenin Sources},
  pdfauthor={Majid Manteghi},
  pdfsubject={Fully coupled multiport power coordinates, scattering, TARC, accepted power, and antenna radiation efficiency},
  pdfkeywords={available power, coupled multiport sources, scattering matrix, TARC, phase-only excitation, radiation efficiency}
}

\begin{document}
\maketitle

\begin{abstract}
A power-normalized scattering representation is developed at a fixed frequency
for a passive linear time-invariant multiport load driven by a fully coupled
multiport Th\'evenin source.  The source is obtained by reducing, at the load reference planes, an
independent-source network whose suppressed internal impedance is passive,
together with an intervening passive matching network.  No diagonal-reference,
uncoupled-source-channel, reciprocity, or commutation assumption is required.  With
$\Rs=\Herm{\Zs}\succ\mathbf0$, completing the square in accepted power
identifies the available-power current and motivates the coordinates
\begin{equation*}
\avec=\frac12\Rs^{-1/2}(\Vvec+\Zs\Ivec),
\qquad
\bvec=\frac12\Rs^{-1/2}(\Vvec-\Zs^H\Ivec).
\end{equation*}
They satisfy
$\|\avec\|_2^2-\|\bvec\|_2^2=\operatorname{Re}\{\Ivec^H\Vvec\}$ and yield
\begin{equation*}
\Smat=\Rs^{-1/2}(\Zload-\Zs^H)
(\Zload+\Zs)^{-1}\Rs^{1/2}.
\end{equation*}
An exact operator identity establishes passivity--contractivity equivalence and
gives excitation-specific, reachable-subspace, and complete conjugate-matching
conditions.  On the physically reachable incident subspace, singular values
characterize the best- and worst-case source-normalized port-reflection TARC,
while a restricted Frobenius norm gives the basis-averaged squared TARC.
Equal-magnitude phase-only control is formulated separately as a
constant-modulus problem, with generator-side constraints mapped through the
coupled source network before power normalization.  For antenna loads and
excitations with $P_{\mathrm{acc}}>0$,
$P_{\mathrm{rad}}/P_{\mathrm{av}}
=\eta_{\mathrm{rad}}(1-\TARC^2)$, so terminal scattering data alone do not
determine radiation efficiency.  When $\Rs$ is singular, finite available
power exists exactly for $\Evec\in\range(\Rs)$, and the construction applies on
the positive-resistance support.
\end{abstract}


\section{Introduction}
\label{sec:introduction}

Scattering parameters are usually introduced through waves referenced to a
scalar impedance or to a diagonal collection of independent port impedances.
That construction is effective when the source channels and their reference
impedances can be treated independently.  A common feed, matching, decoupling,
combining, or distribution network, however, generally produces a full source
impedance matrix at the load reference planes.  The available power is then a
coupled quadratic form, and a diagonal normalization no longer represents the
physical source as a whole.

Complex-reference and power-wave scattering theories have a substantial
history.  Youla considered complex port normalizations \cite{Youla1961};
Rohrer normalized a scattering matrix to a complex $n$-port load network
\cite{Rohrer1965}; and Kurokawa established the widely used diagonal
complex-reference power waves \cite{Kurokawa1965}.  Later work treated real
and complex normalizations, parameter conversion, and the distinction among
traveling waves, pseudo-waves, and power waves
\cite{MarksWilliams1992,ZhuChen1993,Frickey1994,MarksWilliamsFrickey1995}.
Rahola related portwise
power waves to conjugate matching \cite{Rahola2008}.  Separately, Desoer gave
the finite-frequency maximum-power-transfer theorem (MPTT) for $n$-ports,
including singular source resistance and the full class of maximizing loads
\cite{Desoer1973}; passive maximizing loads were further studied by Vidyasagar
\cite{Vidyasagar1974}.  Multiport-antenna efficiency formulations have also
admitted nondiagonal generator impedance matrices \cite{BroydeClavelier2022}.

The contribution here is therefore not priority for full-matrix scattering
normalization.  Rohrer's work is the closest historical antecedent.  The
present purpose and route are different: a physical source and intervening
passive network are first reduced to a Th\'evenin pair at the load planes; the
finite-frequency MPTT current then defines the incident state, its residual
defines the reflected state, and the Hermitian source resistance supplies the
power whitening.  This construction identifies the Euclidean norms directly
with source-available power and its accepted-power deficit.  It also exposes a
qualification that is easily missed: the reduced source-transfer matrix may be
rank deficient.  Consequently, statements about every physically realizable
excitation must be restricted to the incident subspace reachable from the
original source, whereas full-space statements remain valid properties of the
abstract reduced Th\'evenin family.

For antenna loads, two related quantities have both appeared under the name
TARC: a terminal reflected-power ratio and an available-power-referenced
radiated-power deficit.  They coincide for a radiation-lossless antenna but
not when material dissipation is present
\cite{ManteghiRahmatSamii2003,ManteghiRahmatSamii2005,
CapekJelinekMasek2021,CapekJelinek2023,Manteghi2024Comment,
CapekJelinek2024Reply}.  To avoid ambiguity, this paper uses
\emph{source-normalized port-reflection TARC} for the terminal reflection
metric and denotes the total nonradiated-power coefficient separately.

The principal contributions are:
\begin{enumerate}[leftmargin=*]
\item an exact reduction of a coupled source and passive matching network to a
full load-plane Th\'evenin pair, together with a congruence proof that the
suppressed reduced source is passive;
\item MPTT-derived incident and residual power coordinates and the associated
full-matrix scattering representation, without diagonal, reciprocity, or
commutation assumptions;
\item an operator power balance giving passivity--contractivity equivalence and
excitation-specific, reachable-subspace, and complete matching conditions,
including a support-space extension for singular source resistance; and
\item unrestricted and equal-magnitude phase-only TARC extrema, including the
mapping of a physical generator-side phase constraint through the coupled
source network, together with antenna-power consequences that separate
accepted power, radiation efficiency, and material loss and establish
non-identifiability of radiation efficiency from terminal scattering data
alone.
\end{enumerate}

Section~\ref{sec:model} develops the physical reduction and reachable source
subspace.  Sections~\ref{sec:mptt_waves} and \ref{sec:power_balance} construct
the coordinates and establish the operator power balance.  Section~\ref{sec:tarc}
develops TARC and antenna-power consequences, Section~\ref{sec:verification}
provides numerical checks, and Section~\ref{sec:tx_rx} records the transmit and
receive signal-power interpretations.  The appendices prove passivity of the
reduction and treat singular source resistance.

\section{Fully Coupled Source--Matching--Load Model}
\label{sec:model}

All quantities are evaluated at a fixed angular frequency, whose argument is
suppressed.  Voltages and currents are rms phasors, and currents are defined as
entering the passive load.  Thus, the cycle-mean real power entering a load is
\begin{equation}
P=\operatorname{Re}\!\left\{\Ivec^H\Vvec\right\}.
\label{eq:rms_power}
\end{equation}
For a square matrix $\mathbf A$, define
\begin{equation}
\Herm{\mathbf A}\triangleq\frac{1}{2}\left(\mathbf A+\mathbf A^H\right).
\label{eq:Hermitian_part}
\end{equation}
For every nonsingular matrix, the notation
$\mathbf A^{-H}\triangleq(\mathbf A^H)^{-1}$ is used.

\subsection{Physical Interconnection and Th\'evenin Reduction}
\label{subsec:reduction}

Consider an $N$-port Th\'evenin source, a passive $2N$-port matching
network, and an $N$-port load.  After the ideal source voltages are suppressed,
the source impedance is assumed passive,
$\Herm{\mathbf Z_{\mathrm{ss}}}\succeq\mathbf0$.  The matching network may
contain arbitrary coupling among its ports and need not be reciprocal.

\begin{figure}[t]
\centering
\begin{tikzpicture}[
  node distance=10mm and 13mm,
  box/.style={draw, rounded corners, minimum width=31mm, minimum height=10mm,
              align=center, inner sep=2.5mm},
  arr/.style={-{Latex[length=2.2mm]}, thick}
]
\node[box] (src) {$N$-port sources\\$(\mathbf E_{\mathrm s},\mathbf Z_{\mathrm{ss}})$};
\node[box, right=of src] (match) {passive coupled\\$2N$-port network $\mathbf Z_{\mathrm M}$};
\node[box, right=of match] (load) {$N$-port load\\$\Zload$};
\draw[arr] (src) -- (match);
\draw[arr] (match) -- (load);
\node[box, below=13mm of match] (eqsrc) {equivalent source at load planes\\$(\Evec,\Zs)$};
\draw[arr] (match.south) -- node[right, align=left]{port\\reduction} (eqsrc.north);
\draw[arr] (eqsrc.east) -| (load.south);
\end{tikzpicture}
\caption{A fully coupled source and passive $2N$-port matching network reduced
to an equivalent multiport Th\'evenin source at the load reference planes.}
\label{fig:system_model}
\end{figure}

Let $\mathbf V_{\mathrm{in}}$ and $\mathbf I_{\mathrm{in}}$ denote the
source-side voltage and current vectors.  At the output planes, $\Vvec$ is the
load voltage and $\Ivec$ enters the load; therefore, the current entering the
output ports of the matching network is $-\Ivec$.  Partition the matching
network impedance matrix as
\begin{equation}
\begin{bmatrix}
\mathbf V_{\mathrm{in}}\\[1mm]
\Vvec
\end{bmatrix}
=
\begin{bmatrix}
\mathbf Z_{\mathrm{in}} & \mathbf Z_{\mathrm{io}}\\
\mathbf Z_{\mathrm{oi}} & \mathbf Z_{\mathrm{out}}
\end{bmatrix}
\begin{bmatrix}
\mathbf I_{\mathrm{in}}\\[1mm]
-\Ivec
\end{bmatrix}.
\label{eq:matching_partition}
\end{equation}
The source and load equations are
\begin{subequations}
\begin{align}
\mathbf V_{\mathrm{in}}
&=\mathbf E_{\mathrm s}-\mathbf Z_{\mathrm{ss}}\mathbf I_{\mathrm{in}},
\label{eq:raw_source}\\
\Vvec&=\Zload\Ivec.
\label{eq:load_relation}
\end{align}
\end{subequations}
No diagonal structure is assumed for $\mathbf Z_{\mathrm{ss}}$ or for any
block of the matching network.

Define
\begin{equation}
\mathbf A\triangleq\mathbf Z_{\mathrm{in}}+\mathbf Z_{\mathrm{ss}}
\label{eq:A_definition}
\end{equation}
and assume that $\mathbf A$ is nonsingular.  Eliminating
$\mathbf I_{\mathrm{in}}$ gives
\begin{equation}
\Vvec=\Evec-\Zs\Ivec,
\label{eq:equivalent_source_relation}
\end{equation}
where
\begin{subequations}
\label{eq:equivalent_source}
\begin{align}
\Evec
&=\mathbf Z_{\mathrm{oi}}
\left(\mathbf Z_{\mathrm{in}}+\mathbf Z_{\mathrm{ss}}\right)^{-1}
\mathbf E_{\mathrm s},
\label{eq:equivalent_E}\\
\Zs
&=\mathbf Z_{\mathrm{out}}
-\mathbf Z_{\mathrm{oi}}
\left(\mathbf Z_{\mathrm{in}}+\mathbf Z_{\mathrm{ss}}\right)^{-1}
\mathbf Z_{\mathrm{io}}.
\label{eq:equivalent_Zs}
\end{align}
\end{subequations}
Thus $(\Evec,\Zs)$ is the Th\'evenin equivalent of the complete source and
matching network at the load planes.  The second expression is the Schur
complement of $\mathbf A$ in the source-loaded matching matrix
\cite{Zhang2005}.  Even when $\mathbf Z_{\mathrm{ss}}$ is diagonal, $\Zs$ is
generally full.

Define the load-plane source-transfer matrix and its attainable source space by
\begin{subequations}
\begin{align}
\mathbf F&\triangleq
\mathbf Z_{\mathrm{oi}}
(\mathbf Z_{\mathrm{in}}+\mathbf Z_{\mathrm{ss}})^{-1},
&\Evec&=\mathbf F\mathbf E_{\mathrm s},
\label{eq:source_transfer}\\
\mathcal E_{\mathrm r}&\triangleq\range(\mathbf F),
&\mathcal A_{\mathrm r}&\triangleq
\range\!\left(\Rs^{-1/2}\mathbf F\right).
\label{eq:reachable_subspaces}
\end{align}
\end{subequations}
Here the ranges are taken as the modeled independent-source vector
$\mathbf E_{\mathrm s}$ varies over $\mathbb C^N$.  A linear control subspace
may be incorporated by writing $\mathbf E_{\mathrm s}=\mathbf C\mathbf x$
and replacing $\mathbf F$ by $\mathbf F\mathbf C$.  Nonlinear restrictions,
such as equal-magnitude phase-only control, instead restrict the admissible
domain of $\mathbf E_{\mathrm s}$ and are treated separately in
Section~\ref{sec:tarc}.  Let
$q=\rank(\mathbf F)$ and let
$\mathbf U_{\mathrm r}\in\mathbb C^{N\times q}$ have orthonormal columns
spanning $\mathcal A_{\mathrm r}$.  The physically reachable incident states
are
\begin{equation}
\avec=\mathbf U_{\mathrm r}\mathbf c,
\qquad \mathbf c\in\mathbb C^q.
\label{eq:reachable_parameterization}
\end{equation}
If $q<N$, all-excitation matching, extrema, and averages for the original
source must be restricted to $\mathcal A_{\mathrm r}$; if $q=N$, the reachable
space is all of $\mathbb C^N$.  Full-space results remain properties of the
abstract reduced pair $(\Evec,\Zs)$.

Combining \eqref{eq:equivalent_source_relation} and
\eqref{eq:load_relation} gives
\begin{subequations}
\label{eq:terminal_solution}
\begin{align}
\Ivec&=(\Zload+\Zs)^{-1}\Evec,
\label{eq:current_solution}\\
\Vvec&=\Zload(\Zload+\Zs)^{-1}\Evec.
\label{eq:voltage_solution}
\end{align}
\end{subequations}

\subsection{Passivity and Regularity}

Define the source and load resistance matrices
\begin{equation}
\Rs\triangleq\Herm{\Zs},
\qquad
\Rload\triangleq\Herm{\Zload}.
\label{eq:resistance_definitions}
\end{equation}
If the ideal voltage sources are suppressed and the source plus matching
network is passive, Appendix~\ref{app:passivity_reduction} proves
\begin{equation}
\Rs\succeq\mathbf 0.
\label{eq:Rs_psd}
\end{equation}
Likewise, a passive load satisfies $\Rload\succeq\mathbf0$.  Passivity alone
does not imply $\Rs\succ\mathbf0$.  The main development assumes
\begin{equation}
\Rs\succ\mathbf0,
\label{eq:Rs_pd}
\end{equation}
which provides a finite power norm in every source direction and a unique
Hermitian positive-definite square root $\Rs^{1/2}$.  The singular case is
treated in Appendix~\ref{app:singular_Rs}.

Under \eqref{eq:Rs_pd},
\begin{equation}
\Herm{\Zload+\Zs}=\Rload+\Rs\succ\mathbf0.
\label{eq:B_positive}
\end{equation}
This proves that $\Zload+\Zs$ is nonsingular.  Indeed, if
$(\Zload+\Zs)\mathbf x=\mathbf0$ for some $\mathbf x\ne\mathbf0$, then
\begin{equation}
0=\operatorname{Re}\!\left\{\mathbf x^H(\Zload+\Zs)\mathbf x\right\}
=\mathbf x^H(\Rload+\Rs)\mathbf x>0,
\end{equation}
a contradiction.  The same argument applied to $\Zs$ shows that
$\Rs\succ\mathbf0$ already implies that $\Zs$ is nonsingular; no separate
invertibility assumption on $\Zs$ is needed.  Neither reciprocity nor diagonal
structure is assumed.

The model assumes a finite impedance representation and a nonsingular block
$\mathbf A$ at the frequency of interest.  Networks with exact algebraic port
constraints must first be reduced to independent port variables or described
with a compatible hybrid representation.  These are regularity conditions,
not restrictions on the pattern or strength of source coupling.

\section{MPTT-Based Power Coordinates and Scattering Matrix}
\label{sec:mptt_waves}

\subsection{Available-Power Reference State}

For every source-connected terminal state,
\begin{equation}
\Vvec+\Zs\Ivec=\Evec,
\label{eq:source_connected_state}
\end{equation}
the real power entering the load is
\begin{equation}
P_{\mathrm{acc}}
=\operatorname{Re}\!\left\{\Ivec^H\Evec\right\}
-\Ivec^H\Rs\Ivec.
\label{eq:source_power_current}
\end{equation}
The finite-frequency MPTT follows directly by completing the square.

\begin{theorem}[Full-rank MPTT identity]
Under $\Rs\succ\mathbf0$, every source-connected terminal state satisfies
\begin{equation}
\boxed{
P_{\mathrm{acc}}
=\frac14\Evec^H\Rs^{-1}\Evec
-\left(\Ivec-\frac12\Rs^{-1}\Evec\right)^H
\Rs
\left(\Ivec-\frac12\Rs^{-1}\Evec\right).}
\label{eq:full_rank_complete_square}
\end{equation}
Consequently, the unique maximizing current and the available power are
\begin{subequations}
\begin{align}
\mathbf I_{\mathrm{av}}
&=\frac12\Rs^{-1}\Evec,
\label{eq:I_available}\\
P_{\mathrm{av}}
&=\frac14\Evec^H\Rs^{-1}\Evec
=\mathbf I_{\mathrm{av}}^H\Rs\mathbf I_{\mathrm{av}}.
\label{eq:P_available}
\end{align}
\end{subequations}
\end{theorem}

\begin{proof}
Expanding the quadratic term on the right-hand side of
\eqref{eq:full_rank_complete_square} gives
$\operatorname{Re}\{\Ivec^H\Evec\}-\Ivec^H\Rs\Ivec$, which is
\eqref{eq:source_power_current}.  Positive definiteness of $\Rs$ makes the
subtracted quadratic form nonnegative and zero only at
$\Ivec=\mathbf I_{\mathrm{av}}$.
\end{proof}

The complete conjugate termination
\begin{equation}
\mathbf Z_{\mathrm L}=\Zs^H
\label{eq:complete_conjugate_load}
\end{equation}
realizes this current because
$(\Zs+\Zs^H)^{-1}\Evec=\mathbf I_{\mathrm{av}}$.  For a specified nonzero
source vector, complete matrix equality is sufficient but not necessary.  A
well-posed load interconnection realizes the maximizing current if and only if
\begin{equation}
\left(\mathbf Z_{\mathrm L}-\Zs^H\right)\mathbf I_{\mathrm{av}}=\mathbf0.
\label{eq:specified_MPTT_load}
\end{equation}
Indeed,
$\Evec=(\Zs+\Zs^H)\mathbf I_{\mathrm{av}}$, so this condition is equivalent
to $(\mathbf Z_{\mathrm L}+\Zs)\mathbf I_{\mathrm{av}}=\Evec$.
The complete conjugate termination is the unique finite impedance matrix that
produces the maximizing current for every reduced source vector
$\Evec\in\mathbb C^N$ when $\Rs\succ\mathbf0$
\cite{Desoer1973,Vidyasagar1974}.  If only
$\Evec\in\mathcal E_{\mathrm r}$ is physically reachable, the weaker subspace
condition in Section~\ref{sec:power_balance} is sufficient.

\subsection{Incident and Reflected Coordinates}

The MPTT current defines the incident power coordinate
\begin{equation}
\avec\triangleq\Rs^{1/2}\mathbf I_{\mathrm{av}}
=\frac{1}{2}\Rs^{-1/2}\Evec.
\label{eq:a_MPTT}
\end{equation}
Its squared Euclidean norm is exactly the available power:
\begin{equation}
\avec^H\avec=P_{\mathrm{av}}.
\label{eq:a_norm_Pav}
\end{equation}
For the actual load current $\Ivec$, define the reflected coordinate as the
residual from the MPTT current,
\begin{equation}
\bvec\triangleq\Rs^{1/2}
\left(\mathbf I_{\mathrm{av}}-\Ivec\right)
=\avec-\Rs^{1/2}\Ivec.
\label{eq:b_residual}
\end{equation}
Thus $\bvec=\mathbf0$ exactly when the actual interconnection produces the
MPTT current for the specified source state.

Using $\Evec=\Vvec+\Zs\Ivec$, the two coordinates can be written in terminal
variables as
\begin{subequations}
\label{eq:wave_definitions}
\begin{align}
\avec
&=\frac{1}{2}\Rs^{-1/2}(\Vvec+\Zs\Ivec),
\label{eq:a_terminal}\\
\bvec
&=\frac{1}{2}\Rs^{-1/2}(\Vvec-\Zs^H\Ivec).
\label{eq:b_terminal}
\end{align}
\end{subequations}
The order of the matrix factors is essential.  With the rms convention of
Section~\ref{sec:model}, $\avec$ and $\bvec$ have units of
$\sqrt{\mathrm W}$ and $\Smat$ is dimensionless.

\begin{theorem}[Exact power-wave identity]
For arbitrary terminal vectors $\Vvec$ and $\Ivec$, the coordinates in
\eqref{eq:wave_definitions} satisfy
\begin{equation}
\boxed{
\avec^H\avec-\bvec^H\bvec
=\operatorname{Re}\!\left\{\Ivec^H\Vvec\right\}.}
\label{eq:scalar_power_identity}
\end{equation}
\end{theorem}

\begin{proof}
Write $\Zs=\Rs+j\mathbf X_{\mathrm s}$, where
$\mathbf X_{\mathrm s}=\mathbf X_{\mathrm s}^H$.  From
\eqref{eq:wave_definitions},
\begin{equation}
\avec-\bvec=\Rs^{1/2}\Ivec,
\qquad
\avec+\bvec=\Rs^{-1/2}
(\Vvec+j\mathbf X_{\mathrm s}\Ivec).
\end{equation}
Therefore,
\begin{align}
\avec^H\avec-\bvec^H\bvec
&=\operatorname{Re}\!\left\{(\avec+\bvec)^H(\avec-\bvec)\right\}
\nonumber\\
&=\operatorname{Re}\!\left\{\Vvec^H\Ivec
-j\Ivec^H\mathbf X_{\mathrm s}\Ivec\right\}
\nonumber\\
&=\operatorname{Re}\!\left\{\Ivec^H\Vvec\right\},
\end{align}
because $\Ivec^H\mathbf X_{\mathrm s}\Ivec$ is real.
\end{proof}

For a load $\Vvec=\Zload\Ivec$,
\begin{equation}
\bvec^H\bvec=P_{\mathrm{av}}-P_{\mathrm{acc}},
\qquad
P_{\mathrm{acc}}=\Ivec^H\Rload\Ivec.
\label{eq:b_norm_unaccepted}
\end{equation}
Thus, $\bvec^H\bvec$ is exactly the deficit between source-available power
and load-accepted power.  We call this quantity reflected power in the present
MPTT normalization.  For a full $\Rs$, however, the components of $\bvec$
are source-normalized modal coordinates and need not represent independently
measurable backward traveling waves at the physical ports.

\subsection{Scattering Matrix}

Using $\Vvec=\Zload\Ivec$ in \eqref{eq:wave_definitions} gives
\begin{subequations}
\begin{align}
\avec&=\frac{1}{2}\Rs^{-1/2}(\Zload+\Zs)\Ivec,
\label{eq:a_current}\\
\bvec&=\frac{1}{2}\Rs^{-1/2}(\Zload-\Zs^H)\Ivec.
\label{eq:b_current}
\end{align}
\end{subequations}
Eliminating $\Ivec$ yields $\bvec=\Smat\avec$, with
\begin{equation}
\boxed{
\Smat=\Rs^{-1/2}(\Zload-\Zs^H)
(\Zload+\Zs)^{-1}\Rs^{1/2}.}
\label{eq:generalized_S}
\end{equation}
Equivalently,
\begin{equation}
\boxed{
\Smat=\Id_N-2\Rs^{1/2}
(\Zload+\Zs)^{-1}\Rs^{1/2}.}
\label{eq:S_alternate}
\end{equation}
The matrix is determined by the physical source and load impedances, while the
particular source state enters through $\avec$.

For $N=1$, \eqref{eq:wave_definitions} and \eqref{eq:generalized_S} reduce to
\begin{equation}
a=\frac{V+Z_{\mathrm s}I}{2\sqrt{R_{\mathrm s}}},
\quad
b=\frac{V-Z_{\mathrm s}^{*}I}{2\sqrt{R_{\mathrm s}}},
\quad
S=\frac{Z-Z_{\mathrm s}^{*}}{Z+Z_{\mathrm s}}.
\label{eq:one_port_limit}
\end{equation}
For diagonal $\Zs$, the formulas reduce componentwise to Kurokawa's
complex-reference power waves; for a real diagonal $\Zs$, they reduce to the
conventional real-reference scattering formulas.

\subsection{Coordinate Meaning and Unitary Freedom}

When $\Rs$ is diagonal, the components of $\avec$ and $\bvec$ can be associated
with individual physical terminals.  For a full $\Rs$, the whitening
transformation $\Rs^{-1/2}$ generally mixes the terminal coordinates.  The
total norms retain their exact power meaning, but individual components are
source-normalized modal coordinates rather than separately measurable
physical-port powers.

The principal square root supplies a canonical normalization, but the power
coordinates retain a unitary freedom.  For any unitary $\mathbf Q$,
\begin{equation}
\widetilde{\avec}=\mathbf Q\avec,
\qquad
\widetilde{\bvec}=\mathbf Q\bvec,
\qquad
\widetilde{\Smat}=\mathbf Q\Smat\mathbf Q^H.
\label{eq:unitary_change}
\end{equation}
The norms, singular values, and total-power statements are invariant under
\eqref{eq:unitary_change}, whereas individual matrix entries generally are not.

\section{Power Balance, Passivity, and Matching}
\label{sec:power_balance}

Define
\begin{equation}
\mathbf B\triangleq\Zload+\Zs,
\qquad
\mathbf G\triangleq2\mathbf B^{-1}\Rs^{1/2},
\label{eq:B_G}
\end{equation}
so that $\Ivec=\mathbf G\avec$.

\begin{theorem}[Matrix power balance]
The source-normalized scattering matrix satisfies
\begin{align}
\Id_N-\Smat^H\Smat
&=2\Rs^{1/2}\mathbf B^{-H}
(\Zload+\Zload^H)
\mathbf B^{-1}\Rs^{1/2}
\label{eq:matrix_balance_Z}\\
&=4\Rs^{1/2}\mathbf B^{-H}
\Rload\mathbf B^{-1}\Rs^{1/2}
\label{eq:matrix_balance_R}\\
&=\mathbf G^H\Rload\mathbf G.
\label{eq:matrix_balance_G}
\end{align}
\end{theorem}

\begin{proof}
From \eqref{eq:S_alternate},
$\Smat=\Id_N-2\Rs^{1/2}\mathbf B^{-1}\Rs^{1/2}$.  Direct expansion, with
matrix order preserved, gives
\begin{align}
\Id_N-\Smat^H\Smat
&=2\Rs^{1/2}\mathbf B^{-H}
(\mathbf B+\mathbf B^H-2\Rs)
\mathbf B^{-1}\Rs^{1/2}.
\end{align}
Because $\mathbf B+\mathbf B^H-2\Rs=\Zload+\Zload^H$,
\eqref{eq:matrix_balance_Z} follows.  The remaining forms follow from
$\Zload+\Zload^H=2\Rload$ and \eqref{eq:B_G}.
\end{proof}

For every incident state,
\begin{equation}
\avec^H\avec
=\avec^H\Smat^H\Smat\avec+\Ivec^H\Rload\Ivec,
\label{eq:power_ledger}
\end{equation}
which is the operator form of
\begin{equation}
P_{\mathrm{av}}=P_{\mathrm{refl}}+P_{\mathrm{acc}}.
\end{equation}

\subsection{Contractivity and Losslessness}

Since $\mathbf G$ is nonsingular under the main assumptions,
\begin{equation}
\Rload\succeq\mathbf0
\quad\Longleftrightarrow\quad
\Id_N-\Smat^H\Smat\succeq\mathbf0.
\label{eq:passivity_contractivity}
\end{equation}
Thus a passive load produces a contractive scattering matrix,
\begin{equation}
\|\Smat\avec\|_2\leq\|\avec\|_2
\qquad\text{for all }\avec.
\label{eq:contractive}
\end{equation}
If $\Rload\succ\mathbf0$, then $\sigma_{\max}(\Smat)<1$.  If
$\Rload=\mathbf0$, the terminal load accepts no real power and $\Smat$ is
unitary.  A radiation-lossless antenna is not lossless in this terminal-network
sense because radiated power contributes to $\Herm{\Zload}$.

\subsection{Excitation-Specific, Reachable-Subspace, and Complete Matching}

The available power is accepted for a specified nonzero incident state
$\avec_0$ if and only if
\begin{equation}
\Smat\avec_0=\mathbf0.
\label{eq:modal_match}
\end{equation}
With $\Ivec_0=\mathbf G\avec_0$, this is equivalent to
\begin{equation}
(\Zload-\Zs^H)\Ivec_0=\mathbf0.
\label{eq:specified_current_match}
\end{equation}
Matching every state physically reachable from the original source requires
and is equivalent to
\begin{equation}
\boxed{\Smat\mathbf U_{\mathrm r}=\mathbf0,}
\label{eq:reachable_match}
\end{equation}
or, equivalently,
$(\Zload-\Zs^H)\mathbf G\mathbf U_{\mathrm r}=\mathbf0$.  This condition need
not force complete matrix conjugate matching when $q<N$.

Complete matching over the full abstract incident space is stronger:
\begin{equation}
\Smat=\mathbf0
\quad\Longleftrightarrow\quad
\Zload=\Zs^H.
\label{eq:complete_match}
\end{equation}
Thus \eqref{eq:complete_match} is the appropriate all-state condition when
$q=N$ or when a source-independent full-space network property is intended.
Robust matching over an admissible excitation subspace is developed in
Section~\ref{sec:tarc}.

\subsection{Reciprocity}

If the load and reduced source are reciprocal,
\begin{equation}
\Zload^T=\Zload,
\qquad
\Zs^T=\Zs,
\label{eq:reciprocal_impedances}
\end{equation}
then $\Rs$ and its principal square root are real symmetric.  The normalized
matrices
\begin{equation}
\mathbf z=\Rs^{-1/2}\Zload\Rs^{-1/2},
\qquad
\mathbf z_{\mathrm s}=\Rs^{-1/2}\Zs\Rs^{-1/2}
\end{equation}
are symmetric, and
\begin{equation}
\Smat=\Id_N-2(\mathbf z+\mathbf z_{\mathrm s})^{-1}
\end{equation}
is symmetric in the canonical principal-square-root normalization.  A general
complex unitary coordinate change in \eqref{eq:unitary_change} preserves power
and singular values but need not preserve ordinary transpose symmetry.
Reciprocity is not required for the power identity, contractivity, or the
matching statements.

\section{TARC and Maximum Transferred Power}
\label{sec:tarc}

\subsection{TARC as a Total Active Reflection Metric}

For a nonzero source-connected incident state, define the
\emph{source-normalized port-reflection TARC} by
\begin{equation}
\boxed{
\TARC(\avec)
\triangleq
\sqrt{\frac{P_{\mathrm{refl}}}{P_{\mathrm{av}}}}
=\frac{\|\Smat\avec\|_2}{\|\avec\|_2}.}
\label{eq:TARC_definition}
\end{equation}
Here ``total'' refers to the aggregate reflected power over all source-normalized
coordinates, and ``active'' denotes dependence on the simultaneous coherent
multiport excitation; it does not imply an active load.
Its square is the Rayleigh quotient
\begin{equation}
\TARC^2(\avec)
=\frac{\avec^H\Smat^H\Smat\avec}{\avec^H\avec}.
\label{eq:TARC_Rayleigh}
\end{equation}
The accepted-power ratio is exactly
\begin{equation}
\boxed{
\eta_{\mathrm{acc}}(\avec)
\triangleq\frac{P_{\mathrm{acc}}}{P_{\mathrm{av}}}
=1-\TARC^2(\avec).}
\label{eq:eta_acc_TARC}
\end{equation}
This identity is independent of how the accepted real power is partitioned
inside the load.  The numerical value of TARC can of course change when loss
changes the terminal impedance; what TARC does not reveal is whether the
accepted power is useful conversion or dissipation.  In source variables,
\begin{equation}
\eta_{\mathrm{acc}}
=
\frac{4\Evec^H(\Zload+\Zs)^{-H}\Rload
(\Zload+\Zs)^{-1}\Evec}
{\Evec^H\Rs^{-1}\Evec}.
\label{eq:eta_acc_source}
\end{equation}

\subsection{Extrema, Worst-Case Guarantee, and Average Reflection}

Let $\mathcal U\subseteq\mathbb C^N$ be an admissible incident subspace of
dimension $d>0$, and let
$\mathbf U\in\mathbb C^{N\times d}$ have orthonormal columns spanning it.
For the full abstract incident space, take $\mathbf U=\Id_N$ and $d=N$; for
the original physical source, take $\mathbf U=\mathbf U_{\mathrm r}$ and
$d=q$.  Writing $\avec=\mathbf U\mathbf c$ gives
\begin{equation}
\TARC^2(\mathbf U\mathbf c)
=\frac{\mathbf c^H\mathbf U^H\Smat^H\Smat\mathbf U\mathbf c}
{\mathbf c^H\mathbf c}.
\label{eq:subspace_TARC_Rayleigh}
\end{equation}
Hence
\begin{subequations}
\begin{align}
\TARC_{\min}(\mathcal U)
&=\sigma_{\min}(\Smat\mathbf U),
\label{eq:TARC_min}\\
\TARC_{\mathrm{wc}}(\mathcal U)
&=\sigma_{\max}(\Smat\mathbf U).
\label{eq:TARC_max}
\end{align}
\end{subequations}
The corresponding incident states are $\mathbf U\mathbf c$, where
$\mathbf c$ is a right singular vector of $\Smat\mathbf U$.  Therefore,
\begin{equation}
\eta_{\mathrm{acc},\min}(\mathcal U)
=1-\sigma_{\max}^2(\Smat\mathbf U)
\label{eq:guaranteed_transfer}
\end{equation}
is the guaranteed accepted-power fraction over the admissible normalized
source states.  In the full-space case, these reduce to the singular-value
statements for $\Smat$ itself.  Eigenvectors of $\Smat\Smat^H$ describe
outgoing states, not incident excitations.

\paragraph{Equal-magnitude phase-only excitation.}
The singular-value extrema above permit arbitrary complex amplitudes within
an admissible incident subspace.  Phase-dependent TARC under equal-magnitude
port-wave excitation has also been studied directly
\cite{MoradiHonarbakhshEibert2022}.  First suppose that equal magnitude is
imposed on the canonical principal-square-root incident coordinates
of Section~\ref{sec:mptt_waves}.  In the full abstract incident space, write
\begin{equation}
\avec(\boldsymbol\phi)=\alpha\mathbf u(\boldsymbol\phi),
\qquad
\mathbf u(\boldsymbol\phi)=
\begin{bmatrix}
e^{j\phi_1}&e^{j\phi_2}&\cdots&e^{j\phi_N}
\end{bmatrix}^{T},
\qquad |u_n|=1,
\label{eq:phase_only_incident}
\end{equation}
where $\alpha\neq0$.  With
$\mathbf Q\triangleq\Smat^H\Smat$,
\begin{equation}
\boxed{
\TARC_{\mathrm{ph}}^2(\boldsymbol\phi)
=\frac{1}{N}\mathbf u^H(\boldsymbol\phi)
\mathbf Q\mathbf u(\boldsymbol\phi).}
\label{eq:TARC_phase_only}
\end{equation}
A common phase rotation leaves this quotient unchanged, so one phase may be
fixed.  The exact phase-only extrema are
\begin{subequations}
\label{eq:TARC_phase_only_extrema}
\begin{align}
\TARC_{\min,\mathrm{ph}}^2
&=\frac{1}{N}
\min_{\substack{\mathbf u\in\mathbb C^N\\ |u_n|=1}}
\mathbf u^H\mathbf Q\mathbf u,
\label{eq:TARC_phase_only_min}\\
\TARC_{\max,\mathrm{ph}}^2
&=\frac{1}{N}
\max_{\substack{\mathbf u\in\mathbb C^N\\ |u_n|=1}}
\mathbf u^H\mathbf Q\mathbf u.
\label{eq:TARC_phase_only_max}
\end{align}
\end{subequations}
Because the constant-modulus states are a subset of all nonzero incident
states,
\begin{equation}
\boxed{
\sigma_{\min}^2(\Smat)
\leq\TARC_{\min,\mathrm{ph}}^2
\leq\TARC_{\max,\mathrm{ph}}^2
\leq\sigma_{\max}^2(\Smat).}
\label{eq:TARC_phase_only_bounds}
\end{equation}
An endpoint equals its unrestricted singular-value bound only if the
corresponding right-singular subspace contains a constant-modulus vector.  For
a general array, \eqref{eq:TARC_phase_only_extrema} is a nonconvex
constant-modulus quadratic optimization rather than an ordinary eigenvalue
problem.

For a fully coupled physical source, however, equal magnitude is ordinarily a
constraint on the upstream generator channels, not on the whitened components
of $\avec$.  If
\begin{equation}
\mathbf E_{\mathrm s}=E_0\mathbf u,
\qquad |u_n|=1,
\end{equation}
and
\begin{equation}
\mathbf L\triangleq\frac12\Rs^{-1/2}\mathbf F,
\label{eq:source_phase_map}
\end{equation}
then \eqref{eq:source_transfer} and \eqref{eq:a_MPTT} give
$\avec=E_0\mathbf L\mathbf u$.  The physically constrained TARC is therefore
\begin{equation}
\boxed{
\TARC_{\mathrm{src,ph}}^2(\mathbf u)
=
\frac{\mathbf u^H\mathbf L^H\Smat^H\Smat\mathbf L\mathbf u}
{\mathbf u^H\mathbf L^H\mathbf L\mathbf u},
\qquad |u_n|=1,
\quad \mathbf L\mathbf u\neq\mathbf0.}
\label{eq:TARC_source_phase_only}
\end{equation}
Because the denominator in \eqref{eq:TARC_source_phase_only} may vanish at
constant-modulus source vectors, define the lower and upper source-constrained
values by
\begin{subequations}
\label{eq:TARC_source_phase_extrema}
\begin{align}
\underline{\TARC}_{\mathrm{src,ph}}^{\,2}
&\triangleq
\inf_{\substack{\mathbf u\in\mathbb C^N,\ |u_n|=1\\
                  \mathbf L\mathbf u\neq\mathbf0}}
\frac{\mathbf u^H\mathbf L^H\Smat^H\Smat\mathbf L\mathbf u}
     {\mathbf u^H\mathbf L^H\mathbf L\mathbf u},
\label{eq:TARC_source_phase_inf}\\
\overline{\TARC}_{\mathrm{src,ph}}^{\,2}
&\triangleq
\sup_{\substack{\mathbf u\in\mathbb C^N,\ |u_n|=1\\
                  \mathbf L\mathbf u\neq\mathbf0}}
\frac{\mathbf u^H\mathbf L^H\Smat^H\Smat\mathbf L\mathbf u}
     {\mathbf u^H\mathbf L^H\mathbf L\mathbf u}.
\label{eq:TARC_source_phase_sup}
\end{align}
\end{subequations}
Since every feasible incident state lies in $\mathcal A_{\mathrm r}$,
\begin{equation}
\boxed{
\sigma_{\min}^2(\Smat\mathbf U_{\mathrm r})
\leq\underline{\TARC}_{\mathrm{src,ph}}^{\,2}
\leq\overline{\TARC}_{\mathrm{src,ph}}^{\,2}
\leq\sigma_{\max}^2(\Smat\mathbf U_{\mathrm r}),}
\label{eq:TARC_source_phase_bounds}
\end{equation}
provided the feasible set in \eqref{eq:TARC_source_phase_extrema} is nonempty.
If $\min_{|u_n|=1}\|\mathbf L\mathbf u\|_2>0$, the infimum and supremum are
attained and may equivalently be called the minimum and maximum.
Equal-magnitude generator voltages do not generally produce equal-magnitude
incident coordinates after the coupled source network and power whitening.
Moreover, a componentwise constant-modulus constraint is not invariant under
the unitary freedom in \eqref{eq:unitary_change}; it must be imposed in the
physical coordinate system in which the generators and phase shifters reside.

For any orthonormal basis $\{\mathbf u_k\}_{k=1}^{d}$ of $\mathcal U$,
\begin{equation}
\frac{1}{d}\sum_{k=1}^{d}\TARC^2(\mathbf u_k)
=\frac{1}{d}\|\Smat\mathbf U\|_F^2
=\frac{1}{d}\sum_{k=1}^{d}\sigma_k^2(\Smat\mathbf U).
\label{eq:frobenius_average}
\end{equation}
The value is basis independent and also equals the expectation of
$\TARC^2(\mathbf U\mathbf c)$ when $\mathbf c$ is uniform on the complex unit
sphere in $\mathbb C^d$.  In the full canonical incident space, the same mean
squared value is obtained for independent uniformly distributed phases under
equal-magnitude excitation because
$\mathbb E\{\mathbf u\mathbf u^H\}=\Id_N$.  A small average does not preclude
one poorly matched coherent mode.

If a source or matching realization is parameterized by $\mu\in\mathcal M$,
the physically restricted single-frequency minimax problem is
\begin{equation}
\min_{\mu\in\mathcal M}\;
\sigma_{\max}^2\!\left[\Smat(\mu)\mathbf U(\mu)\right],
\label{eq:robust_matching}
\end{equation}
where $\mathbf U(\mu)$ spans the admissible incident subspace.  Setting
$\mathbf U=\Id_N$ gives the full-space conservative problem.  For a fixed
upstream generator, changing the matching network generally changes both
$\Smat(\mu)$ and the reduced incident state and may also change the reachable
subspace; excitation-specific and all-state minimax problems must therefore be
distinguished.

Equation~\eqref{eq:robust_matching} is pointwise in frequency.  Independently
choosing an optimum at every frequency need not define a causal, passive, or
finite-order realizable network.  Broadband optimization must restrict the
underlying network immittance matrices to a single causal passive realization
class---for example, a positive-real impedance realization or an equivalent
bounded-real scattering realization---or regard the pointwise result only as
a bound.

A reference-plane qualification is also essential when the matching network
is varied.  The quantity $P_{\mathrm{av}}(\mu)$ in
\eqref{eq:eta_acc_TARC} is the available power of the reduced load-plane
source; if the intervening network is lossy, that power generally changes with
$\mu$.  Minimizing \eqref{eq:robust_matching} controls mismatch relative to
that reduced source but does not by itself maximize power drawn from a common
upstream generator.  End-to-end optimization across different lossy matching
networks must retain a common upstream power reference or include transducer
loss explicitly.

\subsection{TARC, Accepted Power, and Antenna Radiation Efficiency}
\label{subsec:tarc_efficiency}

The original antenna formulation used radiated power as the transferred-power
term and obtained the feed-port scattering expression under an explicit
lossless-antenna assumption
\cite{ManteghiRahmatSamii2003,ManteghiRahmatSamii2005}.  Under that assumption,
accepted and radiated powers coincide.  For a lossy antenna they do not.  In
the convention adopted here, \eqref{eq:TARC_definition} is explicitly a
port-reflection metric and therefore determines accepted power; the
available-power-referenced radiated deficit is denoted separately below.  This
notation distinguishes the two conventions without assigning either name to
the other quantity
\cite{CapekJelinekMasek2021,CapekJelinek2023,Manteghi2024Comment,
CapekJelinek2024Reply}.

Let the antenna resistance matrix be decomposed as
\begin{equation}
\Rload=\Rrad+\Rloss,
\qquad
\Rrad,\Rloss\succeq\mathbf0,
\label{eq:R_partition}
\end{equation}
where $\Rrad$ represents radiation and $\Rloss$ represents material
dissipation within the chosen load boundary.  Then
\begin{subequations}
\begin{align}
P_{\mathrm{rad}}&=\Ivec^H\Rrad\Ivec,\\
P_{\Omega}&=\Ivec^H\Rloss\Ivec,\\
P_{\mathrm{acc}}&=P_{\mathrm{rad}}+P_{\Omega}.
\end{align}
\label{eq:antenna_power_partition}
\end{subequations}
The anti-Hermitian part of $\Zload$ governs terminal reactive power; reactive
energy storage does not enter this cycle-mean real-power balance.  The two-term
split in \eqref{eq:R_partition} assumes that radiation and material dissipation
exhaust the accepted real power within the chosen antenna boundary.  If other
propagating or dissipative channels are present, their positive-semidefinite
power matrices must be added explicitly.

Using $\Ivec=\mathbf G\avec$, define the accepted, radiation, and material-loss
operators in incident-power coordinates by
\begin{equation}
\Aacc\triangleq\Id_N-\Smat^H\Smat,
\qquad
\Arad\triangleq\mathbf G^H\Rrad\mathbf G,
\qquad
\Aloss\triangleq\mathbf G^H\Rloss\mathbf G.
\label{eq:radiation_operators}
\end{equation}
The matrix power balance gives the exact operator decomposition
\begin{equation}
\boxed{\Aacc=\Arad+\Aloss.}
\label{eq:accepted_radiation_operator_split}
\end{equation}
Consequently,
\begin{equation}
P_{\mathrm{rad}}=\avec^H\Arad\avec,
\qquad
P_{\Omega}=\avec^H\Aloss\avec.
\label{eq:incident_coordinate_rad_loss}
\end{equation}
For an excitation with $P_{\mathrm{acc}}>0$, define the active radiation
efficiency
\begin{equation}
\eta_{\mathrm{rad}}(\avec)
\triangleq
\frac{P_{\mathrm{rad}}}{P_{\mathrm{acc}}}
=
\frac{\avec^H\Arad\avec}
{\avec^H(\Id_N-\Smat^H\Smat)\avec}.
\label{eq:active_radiation_efficiency}
\end{equation}
Because $\Rrad$ and $\Rloss$ need not be proportional, this efficiency can
depend strongly on the coherent multiport excitation.  Define the
available-power-referenced active total efficiency by
\begin{equation}
\boxed{
\eta_{\mathrm{tot}}(\avec)
\triangleq\frac{P_{\mathrm{rad}}}{P_{\mathrm{av}}}
=\eta_{\mathrm{rad}}(\avec)
\left[1-\TARC^2(\avec)\right].}
\label{eq:radiated_power_TARC_efficiency}
\end{equation}
Thus, identifying $P_{\mathrm{rad}}/P_{\mathrm{av}}$ with
$1-\TARC^2$ implicitly assigns unit radiation efficiency to the active
excitation.  Under the port-reflection convention adopted here, radiation
efficiency does not enter TARC; it converts the accepted-power fraction into a
radiated-power fraction.
Consequently, added material loss may reduce TARC by changing the terminal
impedance and increasing the accepted-power fraction while simultaneously
reducing $\eta_{\mathrm{rad}}$.  A lower TARC therefore does not, by itself,
imply a larger radiated-power fraction.

For clarity, denote the available-power-referenced nonradiated fraction by
\begin{equation}
\boxed{
\Gamma_{\mathrm{nr}}^2(\avec)
\triangleq1-\frac{P_{\mathrm{rad}}}{P_{\mathrm{av}}}
=\TARC^2(\avec)+\frac{P_{\Omega}}{P_{\mathrm{av}}}.}
\label{eq:nonradiated_decomposition}
\end{equation}
This quantity includes both terminal reflection and material dissipation.  It
has also been called TARC in parts of the lossy-antenna literature
\cite{CapekJelinekMasek2021,CapekJelinek2023}; the separate symbol here simply
prevents the two power deficits from being conflated.  They coincide when
$\Rloss=\mathbf0$.

\begin{proposition}[Radiation efficiency is not identifiable from terminal
scattering data alone]
For fixed source normalization, the terminal scattering matrix determines
$\Aacc=\Id_N-\Smat^H\Smat$, but it does not determine a unique pair
$(\Arad,\Aloss)$ and therefore does not, in general, determine
$\eta_{\mathrm{rad}}(\avec)$.
\end{proposition}

\begin{proof}
Let $\mathbf K$ be any Hermitian matrix satisfying
$\mathbf0\preceq\mathbf K\preceq\Aacc$.  The decomposition
$\Arad=\mathbf K$ and $\Aloss=\Aacc-\mathbf K$ has the same accepted-power
operator.  Because $\mathbf G$ is nonsingular under the main assumptions,
\begin{equation}
\Rrad=\mathbf G^{-H}\mathbf K\mathbf G^{-1},
\qquad
\Rloss=\mathbf G^{-H}(\Aacc-\mathbf K)\mathbf G^{-1}
\end{equation}
are positive semidefinite and sum to $\Rload$.  These are algebraically
admissible terminal-resistance decompositions and all produce the same terminal
$\Smat$ but generally different radiation efficiencies.  Electromagnetic
realizability may restrict the set of decompositions, but those additional
restrictions are not contained in the terminal scattering matrix.
\end{proof}

The one-port limit gives an immediate counterexample.  A real source
resistance $R_0$ driving either an ideal matched radiator
$(R_{\mathrm{rad}},R_{\Omega})=(R_0,0)$ or a matched dummy load
$(0,R_0)$ has $S=0$ and accepts all available power, while the two radiation
efficiencies are one and zero.  Terminal scattering data therefore provide no
nontrivial efficiency bound without additional electromagnetic or material
information.

For example, choose the embedded far-field normalization so that the radius,
wave-impedance, and rms/time-average factors are absorbed into
$\mathbf f$.  If
$\mathbf f(\widehat{\mathbf r})=\mathbf F_{\!f}(\widehat{\mathbf r})\Ivec$
and the integration is over the unit sphere $\mathbb S^2$, then
$P_{\mathrm{rad}}=\int_{\mathbb S^2}
\|\mathbf f(\widehat{\mathbf r})\|_2^2\,d\Omega$ and
\begin{equation}
\Rrad
=\int_{\mathbb S^2}\mathbf F_{\!f}^H(\widehat{\mathbf r})
\mathbf F_{\!f}(\widehat{\mathbf r})\,d\Omega.
\label{eq:radiation_matrix_from_fields}
\end{equation}
Its off-diagonal entries are embedded-field overlap terms and retain the
radiation-side effects of mutual coupling.  Alternatively, a material-loss
model may provide $\Rloss$, after which $\Rrad=\Rload-\Rloss$.  An augmented
power-conserving scattering description with explicit radiation and loss
channels supplies equivalent factors of $\Arad$ and $\Aloss$; the feed-port
block alone does not \cite{ManteghiRahmatSamii2005}.

For singular $\Rs$, the same construction applies on the finite-power support
space by replacing $\Smat$ with $\mathbf S_+$ and using
$\mathbf G_+=2(\Zload+\Zs)^{-1}\mathbf U_+\boldsymbol\Lambda^{1/2}$.

\section{Numerical Verification}
\label{sec:verification}

The following numerical example verifies the normalization and operator
power balance.  All impedances below are normalized to an arbitrary
positive resistance $Z_0$.  In the reciprocal example, voltages are normalized to $V_0$ and powers
to $V_0^2/Z_0$.  Consider the reciprocal
fully coupled source and passive load
\begin{equation}
\frac{\Zs}{Z_0}=
\begin{bmatrix}
2+j0.6 & 0.45+j0.25\\
0.45+j0.25 & 1.5-j0.3
\end{bmatrix},
\label{eq:example_Zs}
\end{equation}
\begin{equation}
\frac{\Zload}{Z_0}=
\begin{bmatrix}
1.2+j0.4 & 0.35-j0.15\\
0.35-j0.15 & 2.1+j0.2
\end{bmatrix}.
\label{eq:example_Z}
\end{equation}
The eigenvalues of $\Rs/Z_0$ are $1.2352$ and $2.2648$, and those of
$\Rload/Z_0$ are $1.0799$ and $2.2201$; both matrices are positive definite.
Equation~\eqref{eq:generalized_S} gives
\begin{equation}
\Smat\approx
\begin{bmatrix}
-0.13519+j0.36154 & -0.02600+j0.00533\\
-0.02600+j0.00533 & 0.17767-j0.02601
\end{bmatrix}.
\label{eq:example_S}
\end{equation}
Its singular values are
\begin{equation}
\sigma_{\min}=0.17984,
\qquad
\sigma_{\max}=0.38768.
\label{eq:example_singular_values}
\end{equation}
Consequently, the accepted-power fraction lies between $0.84970$ and $0.96766$
over the full abstract unit-norm incident space.  The average squared reflected-power
fraction is
\begin{equation}
\frac{1}{2}\|\Smat\|_F^2=0.09132,
\end{equation}
whereas the worst-case value is
$\sigma_{\max}^2(\Smat)=0.15030$.

For an equal-magnitude canonical incident state
$\avec=\alpha[1\; e^{j\delta}]^T$, let
$\mathbf Q=\Smat^H\Smat=[q_{mn}]$.  The two-port phase-only problem has the
closed form
\begin{equation}
\TARC_{\mathrm{ph}}^2(\delta)
=\frac12\left(q_{11}+q_{22}
+2\operatorname{Re}\{q_{12}e^{j\delta}\}\right),
\label{eq:two_port_phase_TARC}
\end{equation}
and hence
\begin{subequations}
\begin{align}
\TARC_{\min,\mathrm{ph}}^2
&=\frac12\left(q_{11}+q_{22}-2|q_{12}|\right),\\
\TARC_{\max,\mathrm{ph}}^2
&=\frac12\left(q_{11}+q_{22}+2|q_{12}|\right).
\end{align}
\label{eq:two_port_phase_extrema}
\end{subequations}
For \eqref{eq:example_S}, these give
\begin{equation}
\TARC_{\min,\mathrm{ph}}=0.28790,
\qquad
\TARC_{\max,\mathrm{ph}}=0.31584,
\label{eq:example_phase_extrema}
\end{equation}
attained at $\delta_{\min}=94.66^{\circ}$ and
$\delta_{\max}=-85.34^{\circ}$, respectively.  Thus the phase-only interval
lies strictly inside the unrestricted singular-value interval
$[0.17984,0.38768]$ for this example; its extremal singular vectors require
unequal component magnitudes.

For the source vector
\begin{equation}
\Evec=V_0
\begin{bmatrix}
1 & 0.8e^{j40^{\circ}}
\end{bmatrix}^{T},
\label{eq:example_E}
\end{equation}
Table~\ref{tab:example_power} compares direct circuit powers with their
power-wave forms.  Agreement to the displayed precision verifies the MPTT
normalization, the scattering relation, and the scalar power balance; the
circuit value of $P_{\mathrm{refl}}$ is $P_{\mathrm{av}}-P_{\mathrm{acc}}$.

\begin{table}[H]
\caption{Verification for the Fully Coupled Two-Port.  Powers are normalized
to $V_0^2/Z_0$.}
\label{tab:example_power}
\centering
\begin{tabular}{lcc}
\toprule
Quantity & Value & Power-coordinate expression\\
\midrule
$P_{\mathrm{av}}$ & $0.199146$ & $\avec^H\avec$\\
$P_{\mathrm{acc}}$ & $0.180921$ & $\avec^H(\Id_2-\Smat^H\Smat)\avec$\\
$P_{\mathrm{refl}}$ & $0.018226$ & $\bvec^H\bvec$\\
$\TARC(\avec)$ & $0.302520$ & $\|\Smat\avec\|/\|\avec\|$\\
$\eta_{\mathrm{acc}}$ & $0.908481$ & $1-\TARC^2(\avec)$\\
\bottomrule
\end{tabular}
\end{table}

The normalized matrix residual
\begin{equation}
\frac{\|\Id_2-\Smat^H\Smat
-4\Rs^{1/2}(\Zload+\Zs)^{-H}\Rload
(\Zload+\Zs)^{-1}\Rs^{1/2}\|_F}{\|\Id_2\|_F}
\end{equation}
is at machine precision.  Replacing $\Zload$ by $\Zs^H$ makes every element of
$\Smat$ zero, as required by complete multiport conjugate matching.  The
example also shows why a fully coupled source cannot generally be represented
by independent reference impedances: the off-diagonal real part of $\Zs$
contributes directly to the available-power metric and to the whitening
coordinates.

\section{Transmit and Receive Signal-Power Interpretations}
\label{sec:tx_rx}

The same representation applies in either direction once source and load roles
are identified.  In transmission, substitute the reduced transmitter pair
$(\mathbf E_{\mathrm T},\mathbf Z_{\mathrm T})$ for $(\Evec,\Zs)$ and the
antenna impedance $\mathbf Z_{\mathrm A}$ for $\Zload$ in
\eqref{eq:generalized_S}; then $1-\TARC^2$ is the fraction of
transmitter-available power accepted at the antenna planes.

In reception, substitute the illuminated antenna pair
$(\mathbf E_{\mathrm{oc}},\mathbf Z_{\mathrm A})$ as the source and the passive,
well-posed receiver input $\mathbf Z_{\mathrm R}$ as the load.  The same
identity gives the fraction of antenna-available signal power delivered to the
receiver.  This is a signal-power statement only; maximum output SNR also
requires receiver and loss-noise covariance information and is outside the
present scope.

\section{Conclusion}
\label{sec:conclusion}

A power-normalized scattering representation has been derived directly from
the finite-frequency multiport MPTT for a fully coupled Th\'evenin source.  The
incident coordinate is the maximizing current whitened by the Hermitian source
resistance, and the reflected coordinate is its residual from the actual load
current.  Their squared norms are source-available power and the corresponding
accepted-power deficit.  The resulting operator identity proves
passivity--contractivity equivalence without diagonal, reciprocity, or
commutation assumptions.

The distinction among a specified excitation, the physically reachable source
subspace, and the full abstract incident space is essential.  Matching and
worst-case or average TARC metrics must be restricted to the reachable subspace
unless the load-plane source transfer has full rank; complete full-space
matching is equivalent to $\Zload=\Zs^H$.  Under equal-magnitude phase-only
control, the extrema become constant-modulus optimization problems.  When the
constraint is imposed on physical generator channels, it must first be mapped
through the coupled source transfer and the source-resistance whitening, which
generally produces a ratio of quadratic forms rather than a singular-value
problem.

For antennas, port-reflection TARC fixes accepted power but not its partition
between radiation and material loss.  For excitations with
$P_{\mathrm{acc}}>0$, the radiated fraction is
$\eta_{\mathrm{rad}}(1-\TARC^2)$, and radiation efficiency requires field or
loss information beyond the terminal scattering matrix.  When $\Rs$ is
singular, finite available power requires
$\Evec\in\range(\Rs)$ and the same construction applies on the
positive-resistance support.

\appendix
\section{Passivity of the Reduced Source Impedance}
\label{app:passivity_reduction}

With the ideal source voltages suppressed, the source internal impedance adds
to the input block of the matching network.  Using the current orientation of
Section~\ref{subsec:reduction}, define the source-loaded passive matrix
\begin{equation}
\widetilde{\mathbf Z}_{\mathrm M}
=
\begin{bmatrix}
\mathbf Z_{\mathrm{in}}+\mathbf Z_{\mathrm{ss}} & \mathbf Z_{\mathrm{io}}\\
\mathbf Z_{\mathrm{oi}} & \mathbf Z_{\mathrm{out}}
\end{bmatrix}
\equiv
\begin{bmatrix}
\mathbf A_{\mathrm M} & \mathbf B_{\mathrm M}\\
\mathbf C_{\mathrm M} & \mathbf D_{\mathrm M}
\end{bmatrix},
\qquad \mathbf A_{\mathrm M}\ \text{nonsingular}.
\label{eq:app_partition}
\end{equation}
If the suppressed source impedance and matching network are passive, then
$\Herm{\widetilde{\mathbf Z}_{\mathrm M}}\succeq\mathbf0$.  Its Schur complement
is
\begin{equation}
\Zs=\mathbf D_{\mathrm M}
-\mathbf C_{\mathrm M}\mathbf A_{\mathrm M}^{-1}\mathbf B_{\mathrm M}.
\label{eq:app_schur}
\end{equation}
Introduce
\begin{equation}
\mathbf T=
\begin{bmatrix}
-\mathbf A_{\mathrm M}^{-1}\mathbf B_{\mathrm M}\\
\Id_N
\end{bmatrix}.
\label{eq:app_T}
\end{equation}
For an arbitrary current vector entering the output ports, $\mathbf T$ gives
the composite-network current that makes the suppressed source-side voltage
zero.  Consequently,
\begin{equation}
\widetilde{\mathbf Z}_{\mathrm M}\mathbf T
=
\begin{bmatrix}
\mathbf0\\
\Zs
\end{bmatrix},
\qquad
\mathbf T^H\widetilde{\mathbf Z}_{\mathrm M}\mathbf T=\Zs.
\end{equation}
Taking Hermitian parts gives the exact congruence
\begin{equation}
\boxed{
\Herm{\Zs}
=\mathbf T^H\Herm{\widetilde{\mathbf Z}_{\mathrm M}}\mathbf T.}
\label{eq:app_passivity_congruence}
\end{equation}
Consequently,
\begin{equation}
\Herm{\widetilde{\mathbf Z}_{\mathrm M}}\succeq\mathbf0
\quad\Longrightarrow\quad
\Herm{\Zs}\succeq\mathbf0.
\end{equation}
Strict positivity holds if and only if
\begin{equation}
\range(\mathbf T)\cap
\ker\!\left[\Herm{\widetilde{\mathbf Z}_{\mathrm M}}\right]
=\{\mathbf0\}.
\end{equation}
Thus elimination preserves passivity, whereas positive definiteness requires
that no nonzero reduced output mode lie entirely in a lossless null space of
the complete suppressed source-side network.

\section{Singular Source-Resistance Matrices}
\label{app:singular_Rs}

A passive source may have $\Rs\succeq\mathbf0$ with rank $r<N$.  Let
\begin{equation}
\Rs=\mathbf U_+\boldsymbol\Lambda\mathbf U_+^H,
\qquad
\boldsymbol\Lambda\succ\mathbf0,
\label{eq:singular_spectral}
\end{equation}
where $\mathbf U_+\in\mathbb C^{N\times r}$ has orthonormal columns.  Define
\begin{equation}
\mathbf P=\mathbf U_+\mathbf U_+^H,
\quad
\mathbf P_0=\Id_N-\mathbf P,
\quad
\Rs^{\dagger}=\mathbf U_+\boldsymbol\Lambda^{-1}\mathbf U_+^H.
\end{equation}

A finite available power exists only if
\begin{equation}
\mathbf P_0\Evec=\mathbf0,
\qquad\text{equivalently}\qquad
\Evec\in\range(\Rs).
\label{eq:singular_compatibility}
\end{equation}
To see the necessity, write $\Evec=\Evec_++\Evec_0$, with
$\Evec_+=\mathbf P\Evec$ and $\Evec_0=\mathbf P_0\Evec$.  For
$\epsilon>0$, the passive load
\begin{equation}
\mathbf Z_{\mathrm L}(\epsilon)=\Zs^H+\epsilon\mathbf P_0
\end{equation}
produces
\begin{equation}
\mathbf I_{\epsilon}
=\frac12\Rs^{\dagger}\Evec_+
+\frac{1}{\epsilon}\Evec_0.
\end{equation}
Its delivered power contains the term
$\|\Evec_0\|_2^2/\epsilon$, so the supremum over this sequence of passive
loads is unbounded as $\epsilon\rightarrow0^+$ unless
$\Evec_0=\mathbf0$.  No individual member of the sequence delivers literal
infinite power.  The divergence reflects the absence of a current or energy
limit in the ideal Th\'evenin model along an excited zero-resistance direction;
it is not a thermodynamic prohibition of lossless or singular source
impedances.  Under \eqref{eq:singular_compatibility},
\begin{equation}
P_{\mathrm{av}}=\frac14\Evec^H\Rs^{\dagger}\Evec.
\label{eq:singular_Pav}
\end{equation}

For source-connected terminal states satisfying
$\Vvec+\Zs\Ivec=\Evec$ with
$\Evec\in\range(\Rs)$, the accepted power admits the completing-square
identity
\begin{equation}
\boxed{
P_{\mathrm{acc}}
=\frac14\Evec^H\Rs^{\dagger}\Evec
-\left(\Ivec-\frac12\Rs^{\dagger}\Evec\right)^H
\Rs
\left(\Ivec-\frac12\Rs^{\dagger}\Evec\right).}
\label{eq:singular_complete_square}
\end{equation}
Hence the maximum accepted power is $P_{\mathrm{av}}$, and the complete
family of maximizing currents is
\begin{equation}
\boxed{
\Ivec_{\max}=\frac12\Rs^{\dagger}\Evec+\mathbf i_0,
\qquad \mathbf i_0\in\ker(\Rs).}
\label{eq:singular_max_current}
\end{equation}
The maximizing current of the quadratic power functional is therefore generally
nonunique in the singular case.  A particular well-posed load realization may
select only a subset of this family; the support-space power maximum and its
value are independent of that selection.  Under the compatibility condition,
the regularized passive load
$\mathbf Z_{\mathrm L}(\epsilon)=\Zs^H+\epsilon\mathbf P_0$ realizes the
canonical member $\Ivec_{\max}=\tfrac12\Rs^{\dagger}\Evec$ for every
$\epsilon>0$.

When $\Rs$ is singular, passivity alone does not ensure that
$\Zload+\Zs$ is nonsingular.  A useful sufficient condition is
\begin{equation}
\ker(\Rload)\cap\ker(\Rs)=\{\mathbf0\},
\label{eq:singular_nonsingular_sufficient}
\end{equation}
because it implies $\Rload+\Rs\succ\mathbf0$; otherwise well-posedness must be
assumed separately.

For source-connected terminal states satisfying the same compatibility
condition and with $r>0$, define the $r$-component support-space coordinates
\begin{subequations}
\begin{align}
\avec_+
&=\frac12\boldsymbol\Lambda^{-1/2}
\mathbf U_+^H(\Vvec+\Zs\Ivec),
\label{eq:a_plus}\\
\bvec_+
&=\frac12\boldsymbol\Lambda^{-1/2}
\mathbf U_+^H(\Vvec-\Zs^H\Ivec).
\label{eq:b_plus}
\end{align}
\end{subequations}
For these compatible source-connected states,
\begin{equation}
\avec_+^H\avec_+=P_{\mathrm{av}},
\qquad
\avec_+^H\avec_+-\bvec_+^H\bvec_+=P_{\mathrm{acc}}.
\label{eq:singular_support_power_identity}
\end{equation}
When $\Zload+\Zs$ is nonsingular, the support-space scattering matrix is
\begin{align}
\mathbf S_+
&=\boldsymbol\Lambda^{-1/2}\mathbf U_+^H
(\Zload-\Zs^H)(\Zload+\Zs)^{-1}
\mathbf U_+\boldsymbol\Lambda^{1/2}
\label{eq:S_plus}\\
&=\Id_r-2\boldsymbol\Lambda^{1/2}\mathbf U_+^H
(\Zload+\Zs)^{-1}\mathbf U_+
\boldsymbol\Lambda^{1/2}.
\end{align}
It obeys
\begin{align}
\Id_r-\mathbf S_+^H\mathbf S_+
&=4\boldsymbol\Lambda^{1/2}\mathbf U_+^H
(\Zload+\Zs)^{-H}\Rload
(\Zload+\Zs)^{-1}
\mathbf U_+\boldsymbol\Lambda^{1/2}.
\label{eq:singular_operator_balance}
\end{align}
A naive replacement of every inverse by a Moore--Penrose inverse in the full
$N$-dimensional formulas is not sufficient.  Directions in $\ker(\Rs)$ carry
zero source-resistance norm and do not constitute independent finite-power
incident coordinates.  For a physically reduced source, let
$\mathcal C_+\triangleq\ker(\mathbf P_0\mathbf F)$ be the source-control
subspace that satisfies the finite-power compatibility condition.  If
$\mathbf N_+$ has orthonormal columns spanning $\mathcal C_+$, then extrema
and averages must be restricted further to
\begin{equation}
\mathcal A_{\mathrm r,+}
=\range\!\left(
\boldsymbol\Lambda^{-1/2}\mathbf U_+^H\mathbf F\mathbf N_+
\right)
\end{equation}
within the positive-resistance support.

\end{document}